\documentclass[12pt,a4paper]{conference}

\usepackage{fancyhdr}
\usepackage{graphicx,amsmath,amssymb,cite}
\usepackage{multind}

\newcommand{\beq}{\begin{equation}}
\newcommand{\eeq}[1]{\label{#1}\end{equation}}
\newcommand{\eeqn}{\end{equation}}

\newcommand{\beqa}{\begin{eqnarray}}
\newcommand{\eeqa}[1]{\label{#1}\end{eqnarray}}
\newcommand{\eeqan}{\end{eqnarray}}

\let\bar=\overbar

\newcommand{\ie}{{\it i.e.}}
\newcommand{\eg}{{\it e.g.}}

\newcommand{\Dslash}{\not{\hbox{\kern-4pt $D$}}}
\newcommand{\dslash}{\not{\hbox{\kern-2pt $\del$}}}

\newcommand{\msb}{{\bar{\ssstyle M \kern -1pt S}}}

\def \bracket<#1>{\mbox{$\langle {#1}\rangle$}}
\def \brav #1|{\mbox{$\langle {#1}|$}}
\def \ketv #1>{\mbox{$|{#1}\rangle$}} 
\def \mate<#1|#2|#3>{\mbox{$\langle {#1}|\,{#2}\,|{#3}\rangle$}}
\def \mvec #1{\mbox{\boldmath{${#1}$}}}

\begin{document}

\Chapter{Pion-nucleon $P_{33}$ and $P_{11}$ scatterings in the Lippmann-Schwinger approach}
           {Pion-nucleon $P_{33}$ and $P_{11}$ scatterings}{T. Inoue}
\vspace{-6 cm}\includegraphics[width=6 cm]{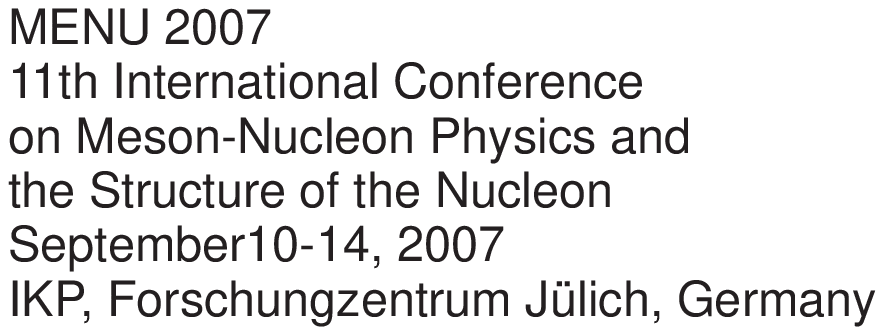}
\vspace{4 cm}

\addcontentsline{toc}{chapter}{{\it T. Inoue}} \label{authorStart}

\begin{raggedright}

{\it Takashi Inoue \footnote{takash-i@sophia.ac.jp}}\index{author}{Inoue, T.}\\
Department of Physics, Sophia University\\
7-1 Kioi-cho, Chiyoda-ku, Tokyo 102-8554, Japan\\
\bigskip\bigskip

\end{raggedright}

\begin{center}
\textbf{Abstract}
\end{center}
We study the pion-nucleon $P_{33}$ and $P_{11}$ scattering,
where the Delta and the Roper resonances are seen, respectively.
We use the Lippmann-Schwinger equation extended to couple to a one-body state,
and investigate nature of these resonances by taking and omitting a one-baryon state into account.
We see validity and puzzle of the standard quark model interpretation, 
for the Delta and the Roper resonances, respectively.

\section{Introduction}

Nowadays, the Delta resonance is well recognized as an appearance of 
a three-constituent-quark(3Q) ground state in the (10,4) representation of 
the flavor SU(3) and spin SU(2). While, before the quark model, 
the resonance was explained as a temporal bound state of pion and nucleon 
produced by attraction due to cross diagram and the centrifugal repulsion. 
The physical resonance in experimental data could be contributed from the both.

Nature of the Roper resonance is interested for many years.
Naive quark model doesn't account for the resonance for the level reverse puzzle,
where the Roper is lighter than the negative parity nucleon $N^*(1535)$.
Graz group claim that the short range part of one-pion-exchange in 3Q system
is enough strong to reverse two masses\cite{Glozman:1995fu}.
While, J\"ulich group argue that the Roper is 
a temporal bound state of nucleon and correlated two pions\cite{Krehl:1999km}.

Then, the purpose of this paper is that
we study the pion-nucleon $P_{33}$ and $P_{11}$ scattering in the Lippmann-Schwinger approach 
with and without including a one-baryon(3Q) state, and see these interpretations.

\index{subject}{the pion-nucleon scattering}
\index{subject}{the Delta resonance}
\index{subject}{the Roper resonance}
\index{subject}{the Lippmann-Schwinger equation}

\section{Formulation}

We solve the Lippmann-Schwinger equation and then T matrix is given by
\begin{equation}
 T_{PP} = M^{-1} V_{PP} \, + \,  M^{-1} V_{PQ} G_Q V_{QP} {M^{-1}}^t
 \quad \mbox{\footnotesize $M = 1 - V_{PP} G_P^{(0)}$}
\end{equation}
\begin{equation}
 G_Q = ( {G_Q^{(0)}}^{-1} - \Sigma_Q )^{-1}  
\qquad
\Sigma_Q = V_{QP} G_P^{(0)} M^{-1} V_{PQ} 
\end{equation}
here the Hilbert space is divided in two with projections $P$ and $Q = 1 - P$
where the P-space is spanned by meson-baryon states, 
while the Q-space is the one-baryon state.
The first term is the T-matrix solved in the P-space only
and is the solution of the ordinary Lippmann-Schwinger equation. 
The second term is the effect of the Q-space \ie\ baryon state,
and is the most important part in this study.
This term is composed of the full propagator of the baryon
including a self-energy due to coupling to pion-nucleon states.

\begin{figure}[t]
\begin{center}
 \includegraphics[width=0.80\textwidth]{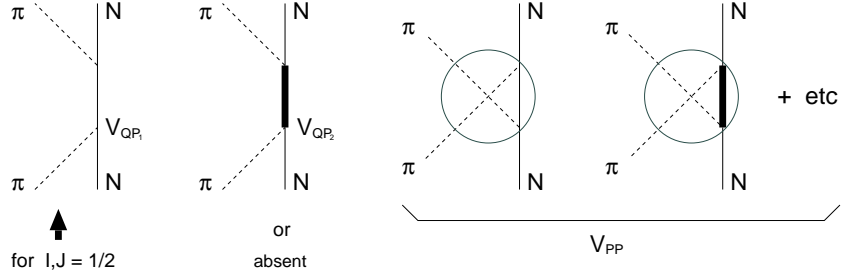}
\caption{Elementary processes taken into account.} 
\label{fig:elepro}
\end{center}
\end{figure}

We consider the elementary processes shown in Fig.\ref{fig:elepro}.
For the $P_{11}$ partial wave, we take the nucleon pole diagram.
The baryon pole diagram, we try both taking and omitting it, as explained.
We include the rest of processes \ie\ the cross-diagram and so on,
as an effective potential in the P-space, $V_{P\!P}$.
We parametrize the matrix elements of these interactions as 
\begin{eqnarray}
 \mate<Q|V_{QP}|\mvec k>
 &=& b^{5/2} \, V_{QP} \ {\mvec k} \cdot \mvec \Sigma \  e^{-\frac{b^2}{2} k^2} 
\\
 \mate<\mvec k'|V_{P\!P}|\mvec k>
 &=& a^5 \, V_{P\!P} \ {\mvec k \cdot \mvec k'} \
          e^{-\frac{a^2}{4} k^2} \, e^{-\frac{a^2}{4} {k'\,}^2} 
\end{eqnarray}
where $\mvec \Sigma$ is the so called transition spin operator 
which is nothing but the spin operator for $\pi NN$ interaction.
Note the (bi-)linear momentum dependence due to P-wave interaction.
The sum over intermediate states is taken by using the Gauss integral method.
We include Gaussian form-factors for simplicity so that the integral converges.
We apply the semi-relativistic prescription.

\section{Result and Discussion}

\begin{figure}[t] 
 \centering
 \includegraphics[width=0.47\textwidth]{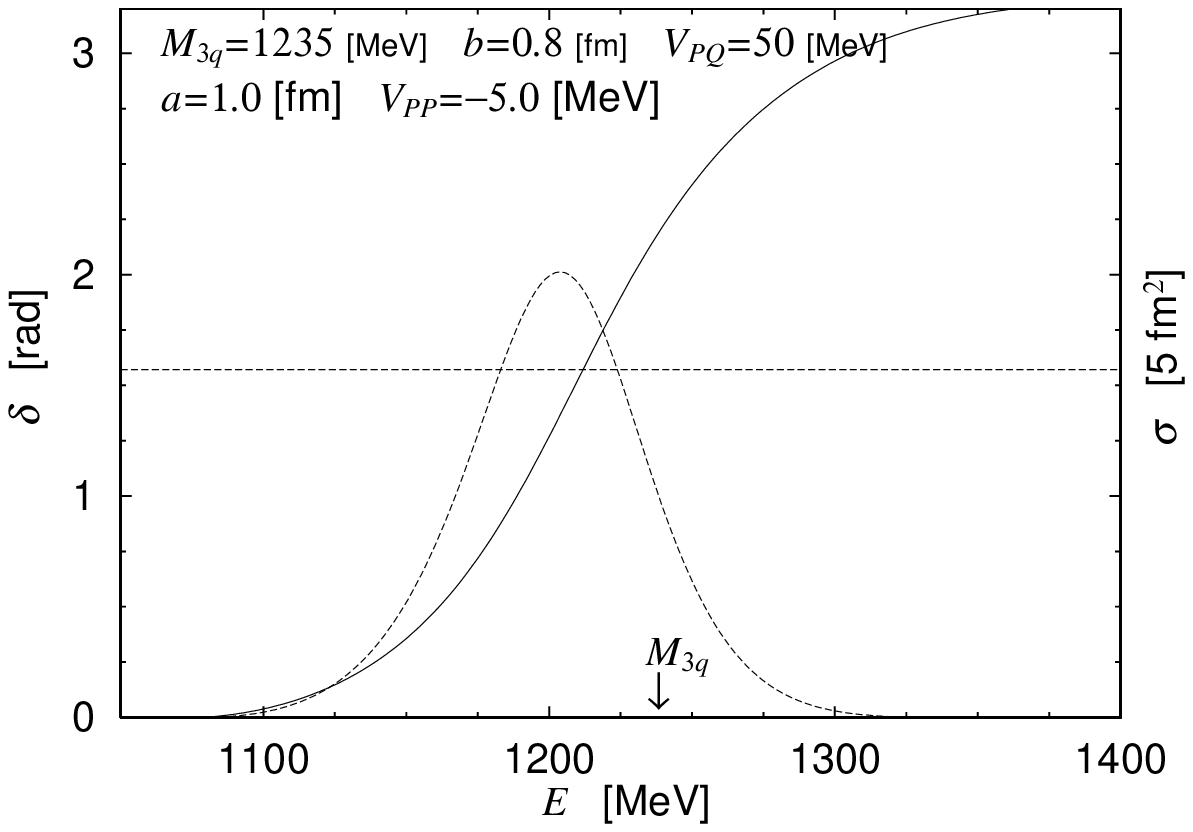}
 \includegraphics[width=0.47\textwidth]{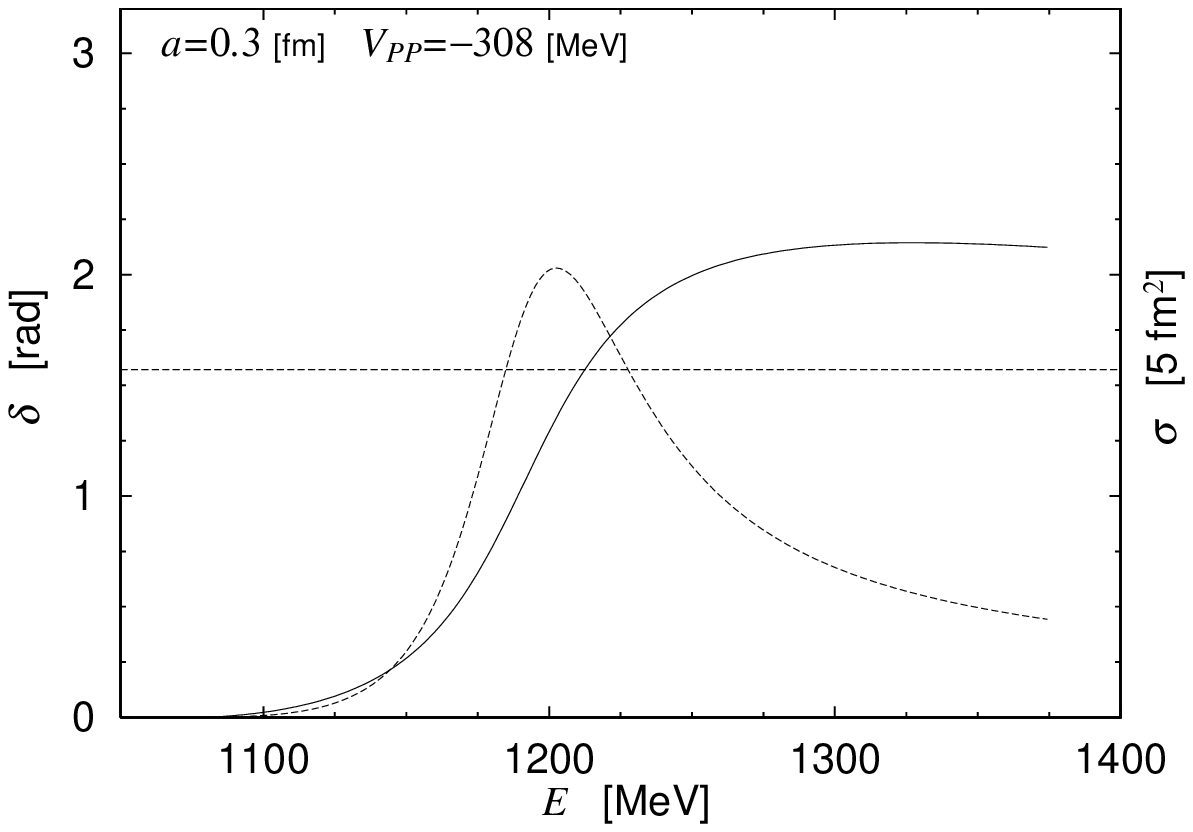}
\caption{Phase-shift and cross-section of $P_{33}$ scattering v.s. the total energy,
         obtained in the model with(left) and without(right) baryon pole.} 
\label{fig:p33}
\end{figure} 

We start with $P_{33}$ partial wave scattering.
Fig.\ref{fig:p33} left shows the phase-shift and cross-section obtained
in the model with a baryon pole, as a function of the total energy.
Entire feature of data is well reproduced by putting a baryon state around 1235[MeV].
Essentially the potential $V_{P\!P}$ is not needed
but a weakly attractive and soft one is added to improved the threshold behavior.
The resonance appears at the almost bare energy of the baryon(slightly below).
This supports the standard quark model scenario.

The right figure of Fig.\ref{fig:p33} shows the results
obtained in the model without a baryon pole.
By assuming a potential with a particular combinations of range and depth,
\eg\ $a=0.3$[fm] and $V_{P\!P}=-308$[MeV], a resonance occur at the right energy.
However, within the present energy independent potential,
the result differ from data at energy above the resonance.
This demonstrate the necessity of a baryon couple to $\pi N$ in $P_{33}$ state.
The above $V_{P\!P}$ should be regarded as an effective potential
including the baryon. 

\begin{figure}[t] 
 \centering
 \includegraphics[width=0.47\textwidth]{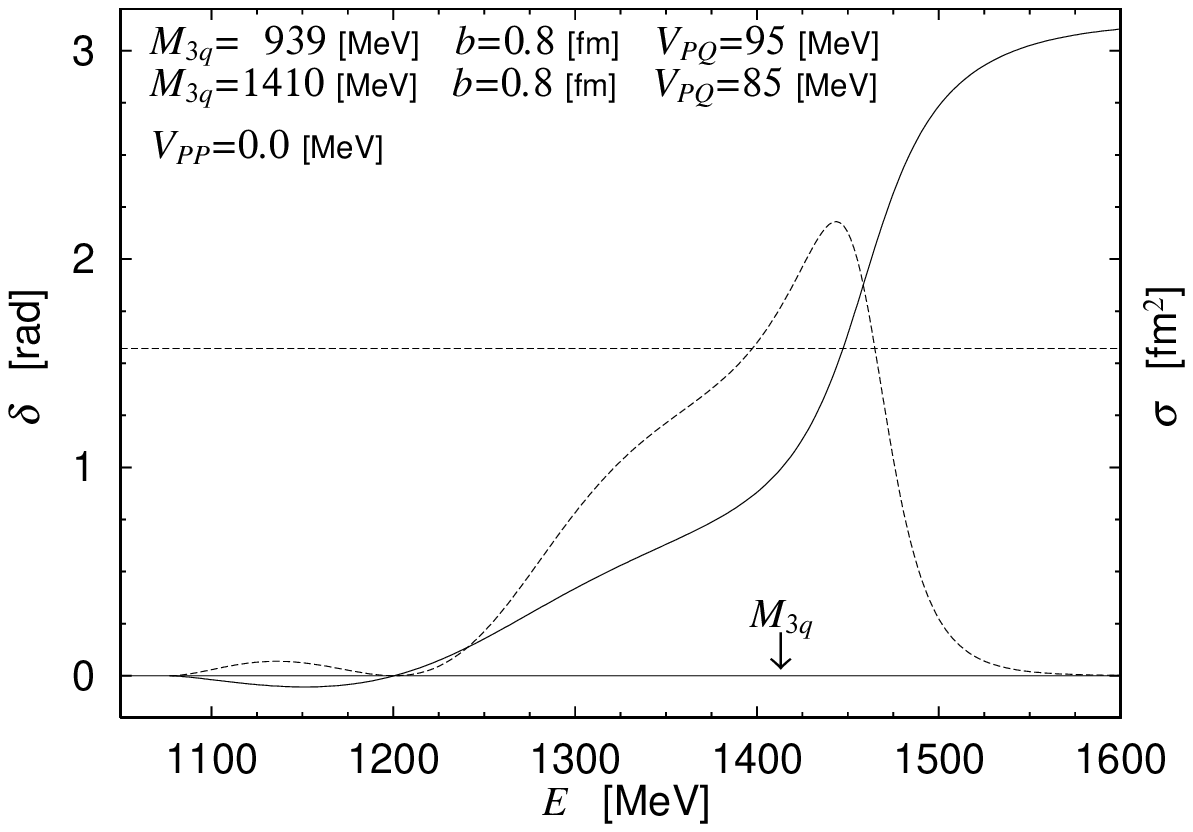}
 \includegraphics[width=0.47\textwidth]{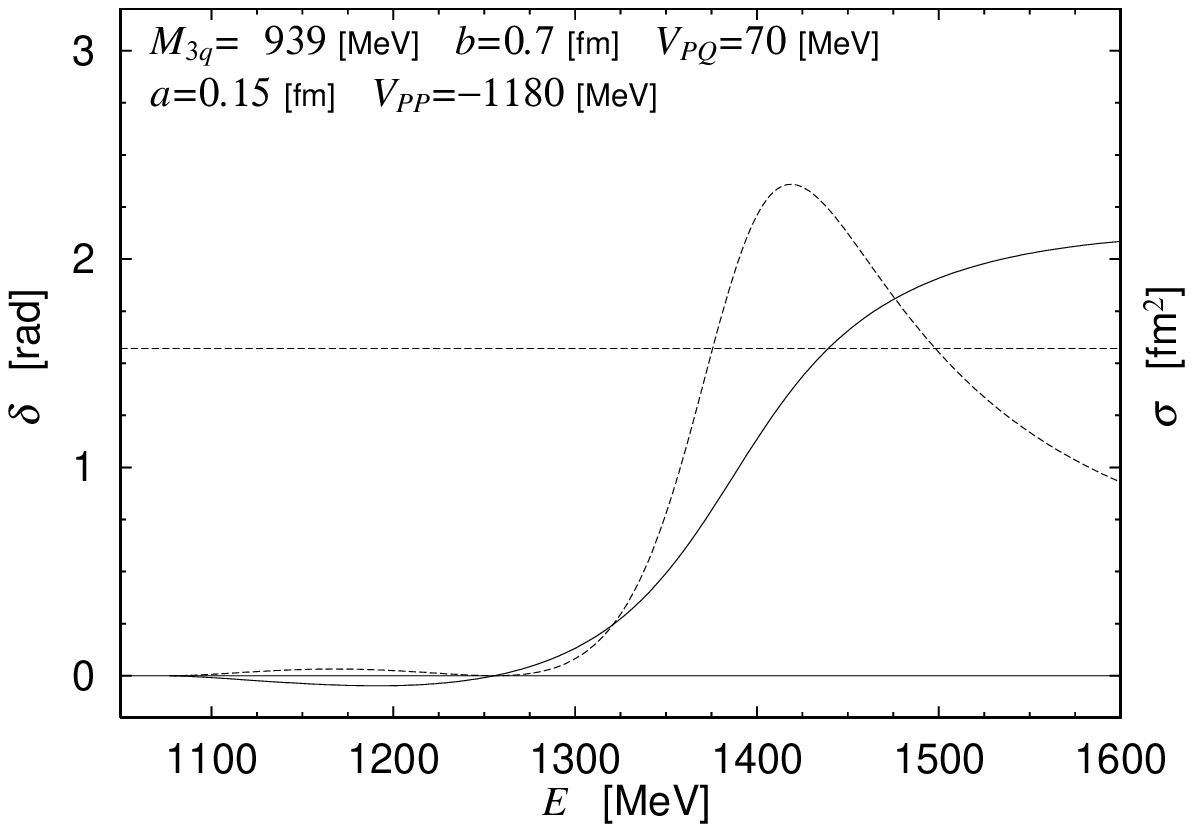}
\caption{Phase-shift and cross-section of $P_{11}$ scattering v.s. the total energy,
         obtained in the model with(left) and without(right) baryon pole.} 
\label{fig:p11}
\end{figure}

We turn to $P_{11}$ partial wave scattering.
Fig.\ref{fig:p11} left shows the result in the model with a baryon pole in addition to nucleon one.
Entire feature of data is well reproduced by putting two baryons at 939[MeV] and 1410[MeV].
Again, no need of $V_{P\!P}$ essentially. 
Note that the bare energy of the baryon is 1410[MeV], 
and the resonance appears at energy slightly above. 
This means that the baryon is shifted to upward.
This result doesn't help the conventional quark model at all. 
We know that the model does not predict a positive parity excited nucleon
around the energy, but around 1550[MeV].
Hence, it must be shifted to downward largely in order to reproduce data.
However, the present result shows that a baryon state will be shifted to upward
by coupling to $\pi N$ scattering state.
After all, the puzzle of Roper resonance in the model, still remains. 

Then, we try to reproduce $P_{11}$ data in absence of a baryon corresponding to the resonance.
The right figure of Fig.\ref{fig:p11} shows the result obtained 
with the nucleon pole and an effective potential.
Again, by assuming a potential with a particular combinations of range and depth,
a resonance occur at the right energy.
This may indicate hadron dynamical origin of the Roper resonance.
In this case, we need a very short range and deep potential,
\eg\ $a=0.15$[fm] and $V_{P\!P}=-1180$[MeV].
We do not have explicit $\pi\pi N$ channel in the present study, 
and the process $\pi N\!\to\!\pi\pi N\!\to\!\pi N$ is regarded as a part of $V_{P\!P}$.
The above deep potential may be originated from the process, 
\eg\ with the strong $\pi\pi$ correlation in the paper by J\"ulich group\cite{Krehl:1999km}. 
This is interesting and must be studied into more details.

\section*{Acknowledgments}
This work is supported by the Grant for Scientific Research No.18042006 
from the Ministry of Education, Culture, Science and Technology, Japan.


%

\end{document}